\documentstyle[11pt,newpasp,twoside,epsf]{article}

\def\edcomment#1{\iffalse\marginpar{\raggedright\sl#1\/}\else\relax\fi}
\reversemarginpar

\begin{document}
\title{Gigahertz-Peaked Spectrum Radio Sources}
\author{M. L. Lister}
\affil{NRAO, 520 Edgemont Road, Charlottesville, VA, USA 22903}

\begin{abstract}
The study of compact AGN which possess convex radio spectra (the gigahertz-peaked
spectrum, or GPS sources) has generated a lot of interest over the
last four decades, since these objects offer a unique opportunity to
probe both the early evolutionary stages of relativistic AGN jets and
their immediate nuclear environments. In this article I trace Ken
Kellermann's early investigations of these sources, which played a
major role in justifying the development of modern-day VLBI
techniques. I describe how our understanding of these AGN has
progressed since Kellermann's early discoveries, and discuss several
ways in which the current classification scheme can be simplified to
reflect intrinsic source characteristics, rather than observer-biased
quantities. Finally, I discuss recent results from the VLBA 2 cm
survey concerning the relativistic jet kinematics of the two-sided
peaked-spectrum sources 4C +12.50 (PKS 1345+125) and OQ 208
(1404+286). 
\end{abstract}

\section{Initial discovery}

The history of peaked-spectrum radio sources can be traced back to Ken
Kellermann's graduate student days at the Owens Valley
Radio Observatory in the early 1960s. At that time, the revised 3C catalog
had just been published (Bennett 1962), and there was great interest
in obtaining accurate broad-band radio spectra of these bright
sources. For his Ph.D. thesis, Kellermann used the variable spacing
interferometer at Owens Valley to obtain flux densities at four
frequencies between 750 and 3200 MHz for 160 objects from the 3C list
and other radio catalogs available at that time. Approximately
half of the sample had straight, steep spectra consistent with
incoherent synchrotron emission from a population of electrons having
a power-law energy distribution.

The remaining radio sources had somewhat curved spectra, and two
unusual objects: CTA 21 (0316+161) and CTA 102 (2230+114), had convex
profiles that peaked around 900 MHz (Kellermann et al. 1962), hence
the origin of the term ``gigahertz-peaked spectrum'' (GPS)
sources. These two CTA objects had been discovered serendipitously by
Dan Harris at Owens Valley several years earlier, and subsequently
generated a great deal of excitement in the Soviet Union as possible
beacons from extraterrestrial civilizations. Kardashev (1964) had
argued that considering the effects of quantum noise and competing
background emission, the most efficient spectral profile for
transmitting radio signals across the galaxy was remarkably similar to
that of CTA 21 and CTA 102.  This coincidence fueled intense
speculation as to their origin. The situation grew more interesting
when Sholomitskii (1965) reported highly unusual flux density
variations in CTA 102 at 32 cm.  This motivated Kellermann (1966) to
look for ``artificial'' features as predicted by Kardashev (1964) in
the spectrum of PKS 1934-638, but none were present. 

Kellermann et al. (1962) had realized early on that the 3C sources
with distinctly curved spectra had the highest brightness
temperatures, but incorrectly attributed this to spectral ageing
processes. Slysh (1963) was subsequently able to fit the curved
spectra of 3C 48 (0134+329) and CTA 21 using formulae he derived for a
synchrotron self-absorbed source, and calculated their angular sizes
to be exceedingly small (140 and 10 milliarcsec, respectively). This
led Shklovsky (1965) to reason, using his spherically expanding
supernova model, that peaked-spectrum sources had to be quite young
and should display substantial flux density variations over time.

At around the same time, Dent (1965) reported significant flux density
changes in 3C 279, and Kellermann \& Pauliny-Toth (1968) began to
regularly monitor a sample of bright AGN at wavelengths ranging from 2
to 40 cm. They soon discovered that flux variability was quite common
in flat-spectrum AGN, but virtually absent in the peaked-spectrum
sources. This posed serious problems for Shklovsky's simple spherical
expansion model for AGN variability, which was later dropped in favor
of more complex shock-in-jet models (e.g., Aller et al. 1985). Several
groups, including Maltby \& Moffet (1965), failed to confirm
Sholomitskii's discovery of variability in CTA 102, although later
low-frequency observations of this source (e.g., Hunstead 1972) showed
flux density variations that were probably associated with
interplanetary scintillation.

\subsection{Early VLBI observations}

By 1966 there was overwhelming indirect evidence that many of the
brightest radio-loud AGN were exceedingly compact.  Both the high
turnover frequencies of the peaked-spectrum sources and light travel
time arguments applied to source variability suggested that some AGN
could only be studied using very long interferometric baselines
exceeding several million wavelengths.  Kellermann (1964) had pointed
out two years earlier that that the peaked-spectrum sources were all
intrinsically very luminous, and therefore could provide a number of
good targets for very long baseline interferometry (VLBI).

These findings provided the scientific motivation for several groups
around the globe to independently pursue the development of VLBI
techniques. Early VLBI observations between Green Bank and Haystack in
1967 at a wavelength of 18 cm (Clark et al. 1968) showed that many AGN
were still unresolved, which underscored the need for even longer
baselines at shorter wavelengths. The following year, Kellermann et
al. (1968) obtained the first inter-continental fringes of several AGN
at 6 cm, between Green Bank and Onsala, Sweden. These fringe
detections gave the first direct measurement of sizes 1 mas or smaller
for several AGN, including the peaked-spectrum source 2134+004.

The direct confirmation of small source sizes using VLBI renewed a
debate launched two years earlier by Kellermann (1966) in his analysis
of the spectrum of PKS 1934-638: was the low-frequency spectral
turnover in peaked-spectrum sources due to free-free absorption (FFA) by thermal
plasma, or synchrotron self-absorption (SSA) from relativistic
electrons? Or was it related to the presence of cold plasma, as
described by the Razin (1960) effect?

Surprisingly, this debate still continues nearly forty years later,
with recent papers discussing either FFA (e.g., Bicknell, Dopita,
\& O'Dea 1997) or SSA models (e.g., Snellen et al. 1999). The main
difficulty is that both models contain enough free parameters to fit
virtually any convex spectrum, albeit with somewhat different
constraints (see O'Dea 1998). Multi-frequency VLBI absorption studies
(e.g., Marr, Taylor \& Crawford 2001; Shaffer, Kellermann, \& Cornwell
1999) may eventually settle this issue, but at present it remains
plausible that both mechanisms play a role in creating the spectral
turnovers in these sources.

\section{Classification schemes}
The standard view of peaked-spectrum sources today is that they are compact,
powerful, young radio sources that reside in gas-rich environments at
the centers of active galaxies (O'Dea 1998). Other typically quoted
properties include low fractional polarization, low apparent jet
speeds, and low flux density variability. However, in reality most
individual peaked-spectrum sources display only some of these characteristics, and
exceptions to the ``classic'' definition are unfortunately the norm.

Part of the problem is that the current definition is based only on the
overall radio spectrum of an AGN, which is often the sum of many
complex, time-variable emitting regions in a relativistic jet.  The
spectral shape can also be heavily influenced by the external
environment, as in the case of FFA. It could be argued, therefore, that
a peaked spectrum by itself  tells us very little about an AGN, except
for the obvious fact that it is dominated by an emitting region of small
angular size.  This suggests that the term ``GPS'' should either be
abandoned entirely, or qualified in more detail so that it can be used
as a true physical classification, and not solely a phenomenological
one. Below I suggest some simple ways that the GPS classification
could be modified so that it favors intrinsic rather than
observer-biased properties.

\subsection{``Masquerading'' blazars}

A first step toward a more definitive classification scheme would be
to eliminate flat-spectrum, variable AGN (i.e., blazars) from the GPS
class. What distinguishes the blazars from GPS sources is their
bright, flat-spectrum cores, but in some blazars the signature of the
core can be occasionally hidden in its integrated radio
spectrum.  Kovalev et al. (2002) 
have shown that the radio spectra of many blazars
change dramatically with time, often showing remarkably GPS-like
convex profiles during flux density outbursts. This can naturally be
explained by the ejection of a new feature in the jet, whose
individual self-absorbed spectrum briefly dominates the total radio
flux. Many of the ``high-frequency peakers'' found by Torkinoski et
al. (2001) appear to fall into this category.

A small fraction of blazars also display persistent features in their
jets that often outshine the flat-spectrum core. These are likely
associated with either stationary shocks or bends in the jet, and can
give rise to a relatively stable, convex spectrum. In most cases,
however, the signature of the flat-spectrum core can be found in
another region of the spectrum, well away from the spectral peak of
the stationary component. A well-known example is the blazar 4C 39.25 (0923+392),
which has a convex spectrum above 1 GHz, but a distinctively flat
spectrum over the region $0.1 < \nu < 1$ GHz. Ironically, CTA 102 also
shows the distinct flat-spectrum and variability of a blazar at
frequencies above 3 GHz, and should be  excluded from the GPS
category. As it turns out, the only reason CTA 102 was originally
called a GPS source is that it was discovered during a relatively
inactive state when its spectrum happened to be convex. 

In order to eliminate these ``masquerading'' blazars from the
peaked-spectrum class, two additional restrictions should be imposed.
First, a bona fide peaked-spectrum source should display a canonical
convex spectrum that a) peaks at {\it any} radio frequency, and b)
maintains this spectrum consistently at all epochs. A canonical peaked
spectrum has been constructed by de Vries et al. (1997), and has
$\alpha > 0.5$ for $\nu < \nu_m$, where $S_\nu \propto \nu^\alpha$ and
$\nu_m$ represents the spectral turnover frequency. In the high
frequency part of the spectrum well past the turnover where the source
is optically thin, the spectral index is steeper than $-0.5$.

\subsection{Compact-steep spectrum sources}

The current classification scheme is also complicated by the nebulous
division between the GPS and compact steep-spectrum (CSS)
sources. These are often grouped together based on their many shared
characteristics (see the review by O'Dea 1998). For example,
when plotted together on the intrinsic size vs. $\nu_m$ plane, they
form a continuous trend having a single slope ($\mathrm{size} \propto
\nu_m^{-0.65}$; O'Dea \& Baum 1997).  The CSS sources originally got
their name since they were unresolved with connected interferometers
($\theta < 1''$), and had overall steep spectra ($\alpha \simeq
-0.7$). However, subsequent low-frequency observations of CSS revealed
that many in fact had spectral turnovers at a few hundred MHz. The
term GPS then became synonymous with sources with $\nu_m \simeq 1$
GHz, and those with lower frequency turnovers remained CSS sources.

This completely unphysical, arbitrary division generates additional
redshift biases in ``complete'' samples of CSS and GPS sources, since
it depends on the {\it observed} turnover frequency.  Some authors
(e.g., Fanti et al. 1985) have used an intrinsic size criterion of 1
kpc to divide CSS and GPS sources, but again this does not correspond
to any physical quantity. Specifically, there is no evidence of a
bi-modality in the observed size distribution. Instead, 1 kpc merely
corresponds to the typical size of a source with a turnover frequency
of 1 GHz.  Furthermore, not all GPS/CSS have measured redshifts,
making some size determinations impossible. To make matters worse,
many intrinsically small GPS sources appear to have associated,
large-scale emission, that is perhaps the result of past, episodic jet
activity in the source (see review by Lara et al. 2002). It is unclear
how to assign a characteristic size to these sources. Finally, there
is again the issue of ``masquerading'' blazars, many of which tend to
have small projected jet sizes due to foreshortening. These sources
can display steep radio spectra if their flat-spectrum core is
under-luminous compared to the jet. This appears to be the case for the
superluminal CSS sources 3C~138 (0518+165), 3C~147 (0538+498), and
3C~216 (0906+430).

In light of these difficulties, the CSS classification should probably
be reserved for physically small ($< 20$ kpc) sources that truly show
no spectral turnover down to the lowest observed frequencies. Some CSS
might very well be foreshortened, but these should be re-classified as
blazars if their viewing angles can be constrained, for example, by
measurements of superluminal motion. Furthermore, any CSS that shows a
MHz-frequency spectral turnover and has a stable, canonical convex
spectrum should be combined with existing GPS sources under the
blanket classification  ``peaked-spectrum radio source''.

\subsection{Compact symmetric objects: a misnomer?}

With the advent of hybrid imaging techniques for VLBI data in the
early 1980's, it became possible to further classify GPS and CSS
sources on the basis of their parsec-scale morphologies. Phillips and
Mutel (1980) popularized the ``compact double'' class on the basis of
VLBI observations of CTD 93 (1607+268) and 3C 395 (1901+319), which
showed bright lobe-like features separated by a few hundred pc. As the
sensitivity of VLBI arrays steadily improved, more and more ``double''
sources were seen to have faint emission that connected the lobes to a
weak core component. The term ``compact double'' was soon abandoned in
favor of ``compact symmetric object'' (CSO)

Although the term CSO is now widely used in the literature, it is
a somewhat misleading description of these sources. First,
the word ``compact'' can have different meanings depending on the
context. VLBI researchers tend to use this term to describe a source
with a large VLBI-to-single dish flux density ratio (i.e. brightness
temperature), whereas most other astronomers would take it to mean a
physically small object. Indeed, most AGN that are classified as CSOs
do satisfy both of these definitions, but a growing number of CSOs
(and other peaked-spectrum sources as well) appear to be associated with kpc-sized
regions of weak, extended emission (Stanghellini et al. 1998).

The ``symmetric'' part of the class definition is also
problematic. Wilkinson et al. (1994) first introduced the term CSO to
describe sources such as 2352+495, which possess a fairly high degree
of morphological symmetry between their jet and counter-jet. However,
the term CSO is now widely applied to {\it any} GPS source that shows
two-sided VLBI jets, regardless of apparent symmetry.  This is
illustrated by many CSO candidates in the  COINS sample of Peck
and Taylor (2000).

The current definition of a CSO, then, relies solely on two
properties: a GPS spectrum, and a two-sided VLBI jet morphology, with
source compactness and symmetry playing insignificant roles. For the
remainder of this article, therefore, I will drop the term ``CSO'' in
favor of ``two-sided peaked-spectrum source''.

\section{Jet kinematics of two-sided peaked-spectrum sources}

Since 1994, Kellermann and the VLBA 2 cm survey collaboration have been
regularly monitoring a large sample of AGN in order to learn more
about the kinematics and evolution of relativistic jets (Zensus et
al. 2002). Although the goals of the program are mainly statistical,
the sample also includes several peaked-spectrum sources which we are
studying in closer detail. Here we report on two particularly
interesting two-sided peaked-spectrum sources: 4C +12.50 (PKS
1345+125) and OQ~208 (1404+286). For the remainder of this paper we
assume an accelerating cosmology with $\Omega_\lambda = 0.7$,
$\Omega_\lambda = 0.3$, and $H_o = 70$ ${\mathrm km \; s^{-1}\;
Mpc^{-1}}$.

\begin{figure}
\plottwo{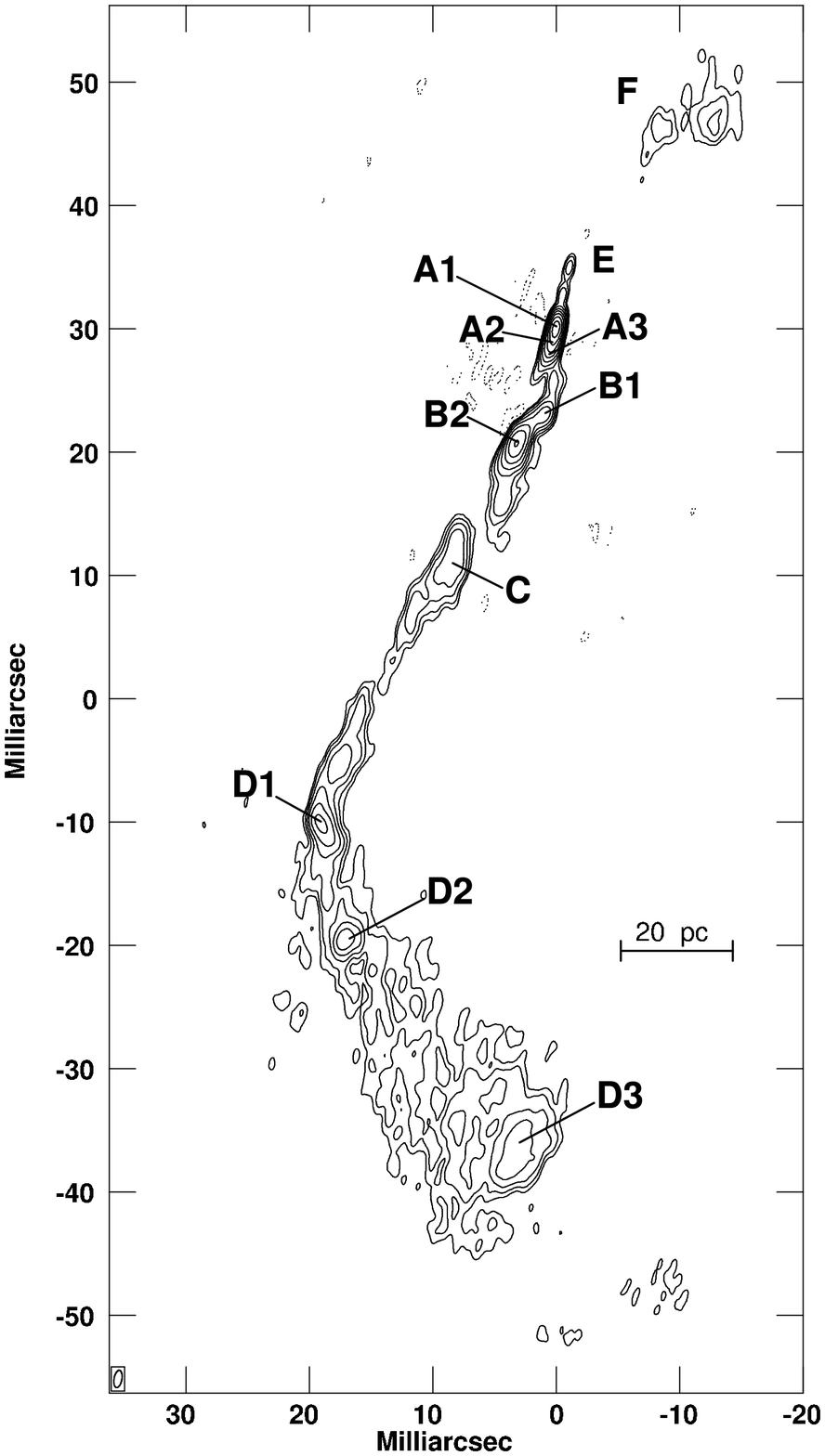}{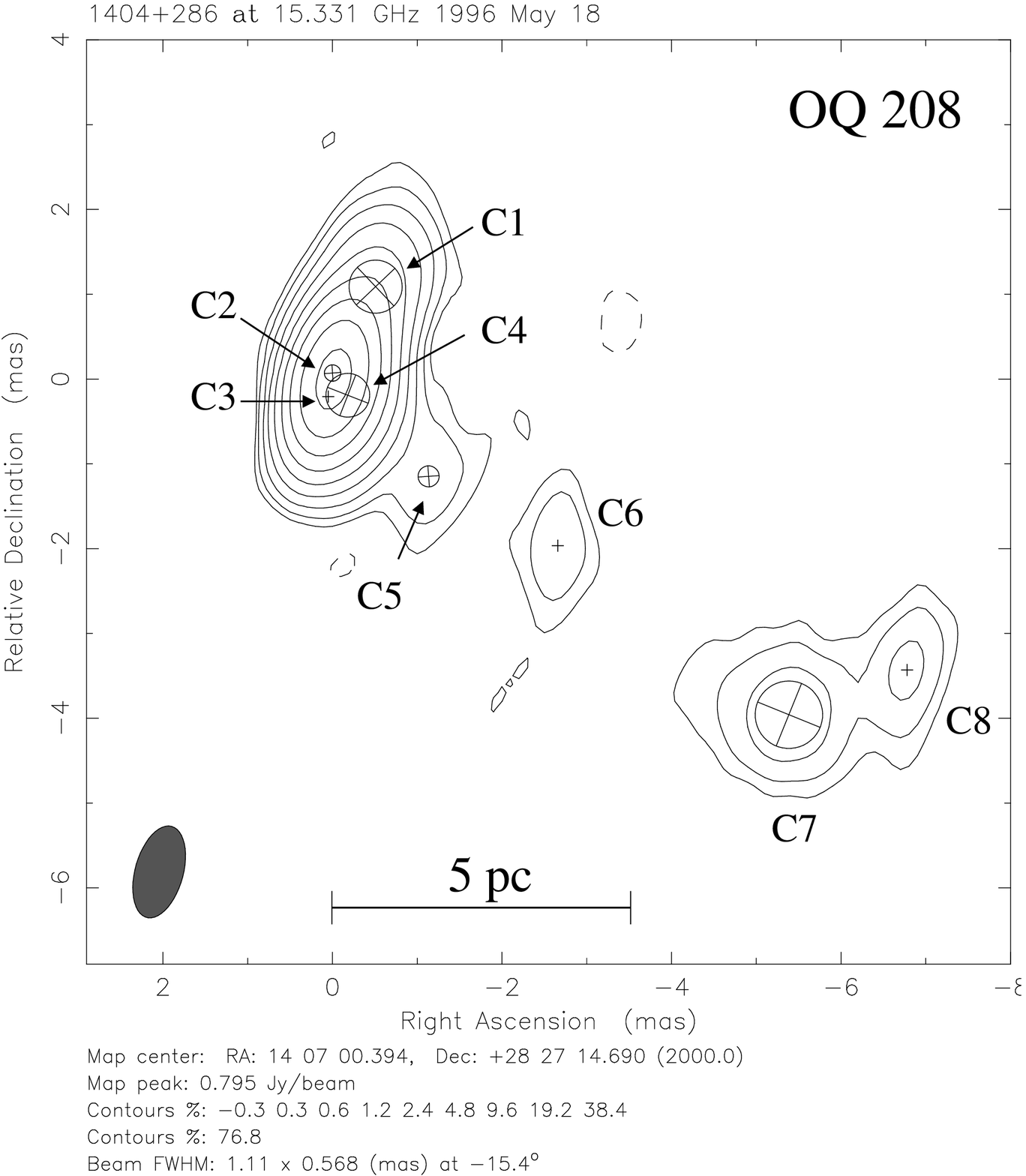}
\caption{VLBA 2 cm images of 4C+12.50 (left panel; from Lister et
al. 2003) and OQ 208 (right panel). }
\end{figure}
\subsection{4C +12.50 }

The powerful radio source 4C +12.50 lies within an ultraluminous
infrared galaxy at $z=0.12$ that is rich in both molecular and atomic
gas.  The host contains two optical nuclei separated by $\sim 3.5$ kpc
that appear to have undergone a recent merger, as evidenced by its
tidal tail structure and distorted isophotes (Gilmore
\& Shaw 1986). Our VLBA observations (Lister et
al. 2003) reveal a well-defined jet structure with a total extent of
$\sim 100$ mas = 220 pc (Fig.~1). The brightest feature in our
full-track VLBA image (A1) has an inverted spectrum and was identified as
the core by Stanghellini et al. (2001). Like most peaked-spectrum sources, the
integrated polarization of 4C +12.50 as measured by the VLA is very
low ($m < 0.2 \%$), however, our analysis has revealed small regions
with extremely high fractional polarization in the southern jet. These
range up to $m = 30\%$ at the major bend (D1) and $m = 60\%$ at the
very tip of the jet (D3). The electric vectors in both regions are
nearly perpendicular to the total intensity contours.

We have detected two components in the southern jet (A2 and A3) that
display substantial apparent motion over our five year sampling
interval. Their mildly superluminal speeds ($\simeq$ 1.2 c) and the
presence of the bright northern counterjet suggest that the jet axes
of 4C +12.50 lie fairly close to the plane of the sky.  We obtained a
good fit to the overall bent morphology of the jet and counter-jet
using a simple constant-wavelength helix of opening angle 23$^\circ$
projected at 82$^\circ$ to the line of sight. Although some helical
jets are the result of material being ejected ballistically from a
precessing nozzle (e.g., SS~433; Hjellming and Johnston 1981), this is
not the case for 4C +12.50, as there is ample evidence for streaming
of material along the curved ridgeline of the jet. These include the
alignment of the polarization gradient and inferred magnetic field
vectors along the jet ridgeline, and the highly polarized flux at
specific points in the jet (i.e., the major bend and southern
tip). These features can naturally be explained as shocks in the flow
that have ordered the magnetic field of the jet.  The helical ridge
line is likely the result of growing Kelvin-Helmholtz instabilities
(e.g., Hardee 1987) that are driven by a small precession of the jet
nozzle. Such a precession might be expected for the black hole of a
galaxy such as this one that has undergone a recent merger event.

The relativistic speeds of the inner jets are significantly higher
than the outer lobe expansion speeds measured for other two-sided
peaked-spectrum sources ($v/c$ $\simeq 0.3$ c; Owsianik, Conway \&
Polatidis 1999). The southern lobe is too diffuse to determine an
accurate advance speed, but if it is typical of other sources, this
implies that a significant amount of jet kinetic energy must be
dissipated over a relatively small distance ($\sim 200$ pc). This is
perhaps being channeled into the large expansion of the jet at D1 or a
terminal shock at D3.

\subsection{OQ~208 }

The radio source OQ~208 is associated with the broad line Seyfert 1
radio galaxy Mrk 668, which like 4C~+12.50, shows signs of tidal
distortion (Stanghellini et al. 1993). Being one of the closest known
peaked-spectrum sources ($z = 0.077$), we are able to study its jet kinematics with
high spatial resolution (1 mas $\simeq 1.5$ pc). Its radio
structure is very compact, with $>99\%$ of its cm flux originating
from a region with a projected diameter of $\sim 12$ pc
(Fig. 1). Aaron (1996) discovered a faint component $\sim 33$ mas SW
of this region at position angle $-107^\circ$. All of the parsec-scale
structure lies within a large diffuse halo of diameter $\sim 30$ kpc
(de Bruyn 1990).

We have assembled ten individual VLBA epochs at 2cm from the VLBA 2 cm
survey and other programs that span the time period 1995-2001. For the
purposes of our proper motion analysis, we have modeled the inner jet
structure using eight discrete circular Gaussian and delta-function
components labeled C1-C8 in Fig. 1. The precise location of the center
of activity in this source is not well known. Our analysis shows the
distance between C1 and C5 is shrinking at a rate of 54 $\mu{\mathrm
as\; y^{-1}}$ ($v/c$ = 0.28), which immediately rules out the
possibility of the core being in the eastern region. Component C6 lies
roughly halfway between the eastern and western ``lobe'' structures,
and is a logical candidate for the core. However, the distance between
C6 and the eastern structure is increasing at rate of $\sim 180$
$\mu{\mathrm as\; y^{-1}}$, while the distance from C6 to the western
structure is {\it decreasing} by 140 $\mu{\mathrm as\; y^{-1}}$. If we
assume there are no inward motions, the core must lie between C5 and
C6, with C6 representing a jet feature that is moving to the SW.

We are unable to determine any individual component velocities without
a stationary reference point, however, C3 and C7 are separating at a
rate of 60 $\pm 4 \; \mu{\mathrm as\; y^{-1}}$.  If OQ~208 has been
steadily expanding at this rate, its parsec-scale emission is only
$\sim 1100 \pm 73$ years old. This is typical of kinematic ages
derived for other two-sided peaked-spectrum sources (Owsianik et
al. 1999).

\section{Summary}

Since Ken Kellermann's initial discoveries regarding
gigahertz-peaked-spectrum sources nearly forty years ago, these
powerful radio sources have greatly increased our understanding of
active galaxies. They not only provided strong incentive for the
development of modern-day VLBI techniques and synchrotron emission
theory in compact radio sources, but are now being used to confirm the
scenario by which AGN jet activity is triggered by galactic
mergers. Unfortunately, the exposure of GPS sources to the larger
astronomical community has been somewhat hindered by a complex and
awkward classification scheme. However, with the small modifications
discussed in this article, it should be possible to implement a scheme
that distinguishes GPS sources on the basis of intrinsic, physical
properties, and not observer-biased ones.

\begin{acknowledgements}
The author would like to thank the members of the VLBA 2 cm
survey collaboration for helpful comments on this manuscript.
\end{acknowledgements}

\end{document}